\documentclass[a4paper,twocolumn,fleqn,superscriptaddress,preprintnumbers]{revtex4}

\usepackage{graphicx,color}
\usepackage{amsmath,amssymb,amsfonts}

\newcommand{\be}{\begin{equation}}
\newcommand{\ee}{\end{equation}}

\newcommand{\bea}{\begin{eqnarray}}
\newcommand{\eea}{\end{eqnarray}}

\newcommand{\ba}{\begin{eqnarray}}
\newcommand{\ea}{\end{eqnarray}}

\def\be{\begin{equation}}
\def\ee{\end{equation}}
\def\beq{\begin{eqnarray}}
\def\eeq{\end{eqnarray}}

\begin{document}

\input amssym.def
\input amssym.tex

\title{Universality in holographic entropy production}

\preprint{BI-TP 2014/13, HIP-2013-19/TH, NIKHEF-2014-12, OUTP-14-09p, TUW-14-06}

\author{Ville Ker\"{a}nen\footnote{vkeranen1@gmail.com}}
\affiliation{Rudolf Peierls Centre for Theoretical Physics, University of Oxford,\\ 1 Keble Road,
Oxford OX1 3NP, United Kingdom}
\author{Hiromichi Nishimura\footnote{nishimura@physik.uni-bielefeld.de}}
\affiliation{Faculty of Physics, Bielefeld University,
33615 Bielefeld, Germany}
\author{Stefan Stricker\footnote{stricker@hep.itp.tuwien.ac.at}}
\affiliation{Institute of Theoretical Physics, Technical University of Vienna,\\ Wiedner Hauptstr.~8-10, 1040 Vienna, Austria}
\author{Olli Taanila\footnote{olli.taanila@iki.fi}}
\affiliation{Nikhef, Science Park 105, 1098 XG Amsterdam, The Netherlands}
\author{Aleksi Vuorinen\footnote{aleksi.vuorinen@helsinki.fi}}
\affiliation{Department of Physics and Helsinki Institute of Physics, P.O.~Box 64, 00014 University of Helsinki, Finland}

\begin{abstract}
We consider the time evolution of two entropy-like quantities, the holographic entanglement entropy and causal holographic information, in a model of holographic thermalization dual to the gravitational collapse of a thin planar shell. Unlike earlier calculations valid in different limits, we perform a full treatment of the dynamics of the system, varying both the shell's equation of state and initial position. In all cases considered, we find that between an early period related to the acceleration of the shell and a late epoch of saturation towards the thermal limit, the entanglement entropy exhibits universal linear growth in time in accordance with the prediction of Liu and Suh. As intermediate steps of our analysis, we explicitly construct a coordinate system continuous at the location of an infinitely thin shell and derive matching conditions for geodesics and extremal surfaces traversing this region.
\end{abstract}

\maketitle

{\em Introduction.}
The equilibration dynamics of strongly coupled systems is an active topic of research, motivated equally by studies of the thermalization process of heavy ion collisions and quantum quenches in condensed matter systems (see e.g.~\cite{DeWolfe:2013cua} for a review). Approaching the problem in the cleanest setup possible --- ${\mathcal N}=4$ Super Yang-Mills (SYM) theory in the limit of large $N_c$ and 't Hooft coupling $\lambda$ --- it may be formulated as follows: Given various physically motivated initial conditions, how does the gravitational system evolve towards its final state involving a black hole in AdS$_5$ space-time? And what are the implications of this gravitational dynamics on the dual field theory side; in particular, how do different physical quantities behave during the equilibration process?

In the context of heavy ion physics, recent years have witnessed remarkable progress in the holographic description of the collision. This includes extensive work on colliding shock waves in strongly coupled ${\mathcal N}=4$ SYM theory \cite{Chesler:2010bi, Wu:2011yd, Casalderrey-Solana:2013aba, vanderSchee:2012qj, vanderSchee:2013pia}, equilibration in inhomogeneous and anisotropic systems \cite{Chesler:2008hg,Heller:2012km,Chesler:2012zk,Heller:2013oxa,Balasubramanian:2013rva}, and even studies relaxing the assumptions of infinite 't Hooft coupling \cite{Steineder:2012si, Steineder:2013ana,Stricker:2013lma, Baron:2013cya} and conformal invariance \cite{Craps:2013iaa}. Typical quantities considered in these setups are expectation values of different components of the energy momentum tensor as well as two-point functions related to transport or photon production \cite{Baier:2012ax,Baier:2012tc}. So far, more complicated observables have been considered only in the simplest setups, such as the quasistatic or Vaidya limits of a gravitationally collapsing planar shell moving either arbitrarily slowly or at the speed of light, respectively (see \cite{Danielsson1,Danielsson2,LinShuryakIII} for more details of this model). Some progress has, however, also been made in the evaluation of Green's functions and geometric probes in a more generic out-of-equilibrium setting \cite{CaronHuot:2011dr,Erdmenger:2012xu,Baron:2012fv}.

On the condensed matter side, the very same model involving a collapsing shell has been applied to the study of spatially uniform quantum quenches, with the most relevant observables being various entropy-like quantities that provide information about the mixing of a system of finite extent with its surroundings. Examples of such calculations include computations of the time evolution of the entanglement entropy (HEE) \cite{AbajoArrastia:2010yt,Albash:2010mv,Baron:2012fv,Liu:2013qca,Abajo-Arrastia:2014fma,Balasubramanian:2011ur} and the causal holographic information (CHI) \cite{Hubeny:2012wa,Hubeny:2013hz}, which follow a line of work initiated in \cite{Ryu:2006bv,Hubeny:2007xt}. These calculations typically reduce to determining the area of some extremal surface, and thus represent purely geometric probes of the field theory system, only sensitive to the metric of the dual space-time. So far, they have however, too, been mostly considered only for lightlike shell trajectories.

Owing to the ubiquity of the collapsing shell model in out-of-equilibrium holography, it is clearly important to develop machinery for solving gravitational problems in this background. In particular, it would be useful to be able to proceed beyond the usual quasistatic and Vaidya limits in the determination of both geometric probes and two-point functions. This constitutes our long-term goal, towards which the paper at hand takes the first steps. Here, our plan is the following: First, we study the dynamics of shells that are described by various equations of state (EoS) and released from rest at a fixed radial coordinate $r_0$. This way we can study a large class of collapsing space-times, characterized by different shell trajectories and corresponding to different initial conditions in the dual field theory. Then, we prepare for the determination of extremal surfaces by deriving matching conditions for geodesics traversing the shell, and as a byproduct construct a coordinate system that is continuous at the location of an infinitely thin massive shell. Finally, we apply these results to the determination of two entropy-like quantities, the HEE and CHI, of which the latter has recently been argued to be equal to a coarse grained entropy \cite{Kelly:2013aja}. The dependence of our results on the parameters of the setup is in addition examined, and the key results compared to previous calculations in the Vaidya metric.

{\em Dynamics of the shell.}
We consider a space-time with a negative cosmological constant, whose only matter content is a thin massive shell with an energy momentum tensor proportional to a delta function in the radial coordinate. To this end, we assume that the metric inside ($-$) and outside ($+$) the shell takes the form
\beq
ds^2=-f_{\pm}(r)dt_{\pm}^2+\frac{dr^2}{f_{\pm}(r)}+r^2d\textbf{x}^2 \, ,
\eeq
where the two functions read for the Poincar\'{e} patch of $4+1$ -dimensional AdS spacetime (whose curvature radius $L$ has been set to unity)
\begin{equation}
f_-(r) = r^2 \quad \text{and} \quad f_+(r) = r^2 - \frac{m}{r^2} \, .
\end{equation}
The $r$-coordinate is continuous, and the shell resides at some (time-dependent) value of it, which we shall call $r_s$. The time coordinate, however, is not continuous, and thus there exists a non-trivial relation between the inside and outside times, $t_-$ and $t_+$. In what follows, we furthermore work in units in which $m$ (and thus the Schwarzschild radius of the emerging black hole $r_h$) has been set to unity; this means that we are measuring distances and times in the field theory in units of $1/(\pi\, T_\text{final})$, where $T_\text{final}$ is the temperature of the final equilibrium state.

The shell's equation of motion (EoM) is obtained using the Israel junction conditions \cite{Israel:1966rt} to relate the extrinsic curvatures inside and outside the shell to its energy momentum content. Due to the symmetries of the setup, there are only two independent components of the energy momentum tensor, $T_{xx}=T_{yy}=T_{zz}$ and $T_{\tau\tau}$, where $\tau$ refers to the proper time of the shell. Ideally, the shell would be created by momentarily turning on a source term in the boundary theory \cite{Wu:2012rib,Baier:2013gsa}, in which case the relation between the two components of $T_{\mu\nu}$, i.e.~the EoS of the shell, would be known by construction. For the purposes of our calculation, we will instead simply assume a linear relation between the pressure and energy density of the system, $p=c \epsilon$, allowing us to solve the form of the energy momentum tensor as $T_{\tau\tau}\sim\ r^{-3(1+c)}$.

\begin{figure}[t]
\begin{center}
\includegraphics[scale=0.45]{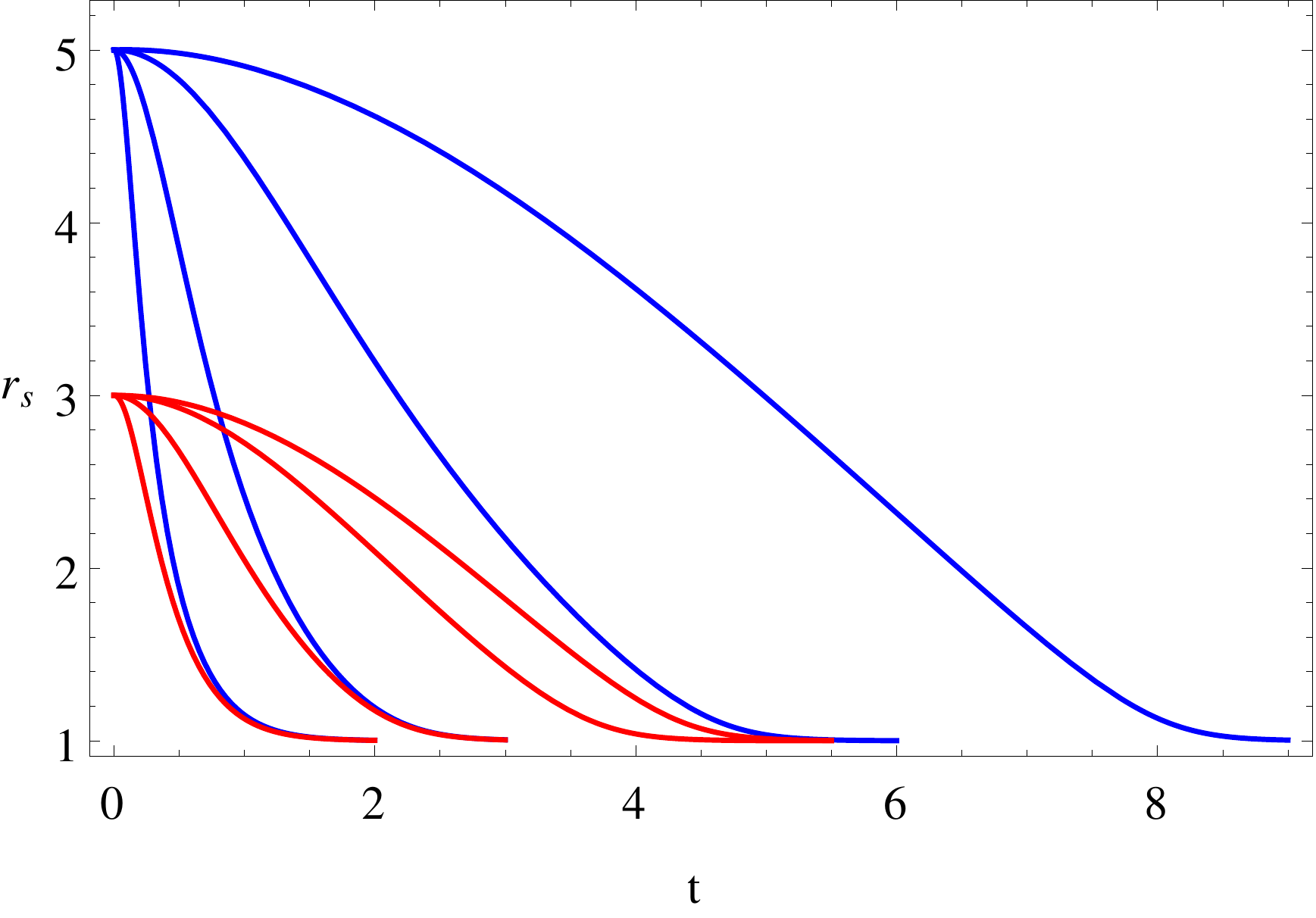}
\caption{\label{fig:traj} 
Two families of shell trajectories that start off from $r_0=3$ and 5 and both comprise of four curves corresponding to EoS parameters $c=0$, 0.3, 0.33 and 1/3 (from left to right). It is interesting to note that the time scale of the gravitational collapse dramatically increases as $c$ approaches the value 1/3.}

\end{center}
\end{figure}

A straightforward calculation now gives as the EoM of the shell
\begin{align}
&\dot{r}_s^2=\frac{M^2}{4r_s^{4+6c}}-\frac{f_-+f_+}{2}+\frac{(f_--f_+)^2r_s^{4+6c}}{4M^2},\nonumber
\\
&\dot{t}_{s\pm}=\frac{\sqrt{f_{\pm}+\dot{r}_s^2}}{f_{\pm}},\label{eq:shelleom1}
\end{align}
where $t_s$ denotes the coordinate time of the shell. The dot on the other hand stands for a derivative with respect to the proper time $\tau$ of the shell, while $M$ is an integration constant related to its energy density. In fig.~\ref{fig:traj} and the text below, we summarize the main features of the solutions to these equations, while a more detailed analysis, including the case of multiple collapsing shells, will be presented in a later publication. 

If the EoS parameter $c$ is within the range $-1\le c < 1/3$, it is easy to see from eq.~(\ref{eq:shelleom1}) that there exists a maximal value of $r_s$, where the shell can reside. We denote this quantity by $r_0$ and call it the turning point of the shell: This can be considered as the location where the shell is released from rest. If its value is taken to infinity (towards the boundary) by tuning $M$ while keeping the mass of the final black hole fixed, the shell trajectory approaches a null geodesic near the boundary. In the interval considered here, this happens irrespective of the value of the EoS parameter $c$, and reduces the space-time to the Vaidya limit.

Close to the conformal value $c=1/3$, in which the trace of the shell's energy momentum tensor vanishes, the near boundary behavior of the trajectory changes qualitatively: In this limit, the shell can either turn around at $r_s=r_0$ or escape to the boundary. Solving the EoM (\ref{eq:shelleom1}) in the limit of a large $r_0$, one finds
\beq
r_s(t_{\pm})=r_0-\frac{1}{r_0}t_{\pm}^2+O(r_0^{-8}),
\eeq
which tells us that the time scale that it takes the shell to move an order one distance is $t\sim \sqrt{r_0}$. Thus, the time it takes for a shell with $c=1/3$ to collapse from $r=r_0$ becomes arbitrarily long as $r_0\rightarrow\infty$.

For the late time behavior of the trajectories, we find universal behavior as a function of $t_+$, 
\beq
r_s(t_+)\approx 1 + C e^{-4t_+},
\eeq
which can be recognized as a null geodesic. Thus, from the point of view of a static observer outside the shell, the shell appears to approach the speed of light near the horizon independent of the value of $c$. 

\begin{figure}[t]
\begin{center}
\includegraphics[scale=1.12]{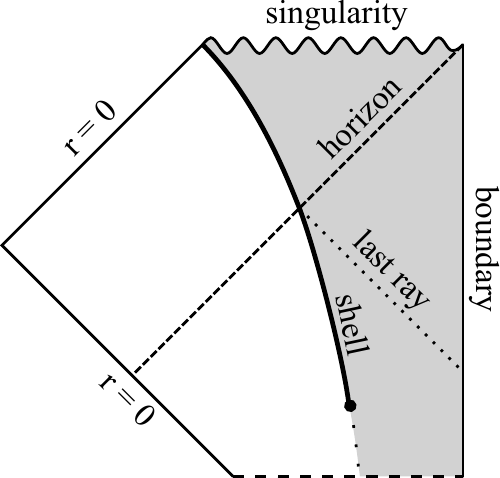}
\caption{\label{fig:penrose} 
Sketch of the Penrose diagram of the collapsing shell space-time, with the past infinity not displayed. Here, the shaded grey region denotes the part of the space-time described by AdS-Schwarzschild metric, while the black dot marks the time and radial location, at which the shell is released from rest.}
\end{center}
\end{figure}

\begin{figure}[t]
\begin{center}
\includegraphics[scale=0.45]{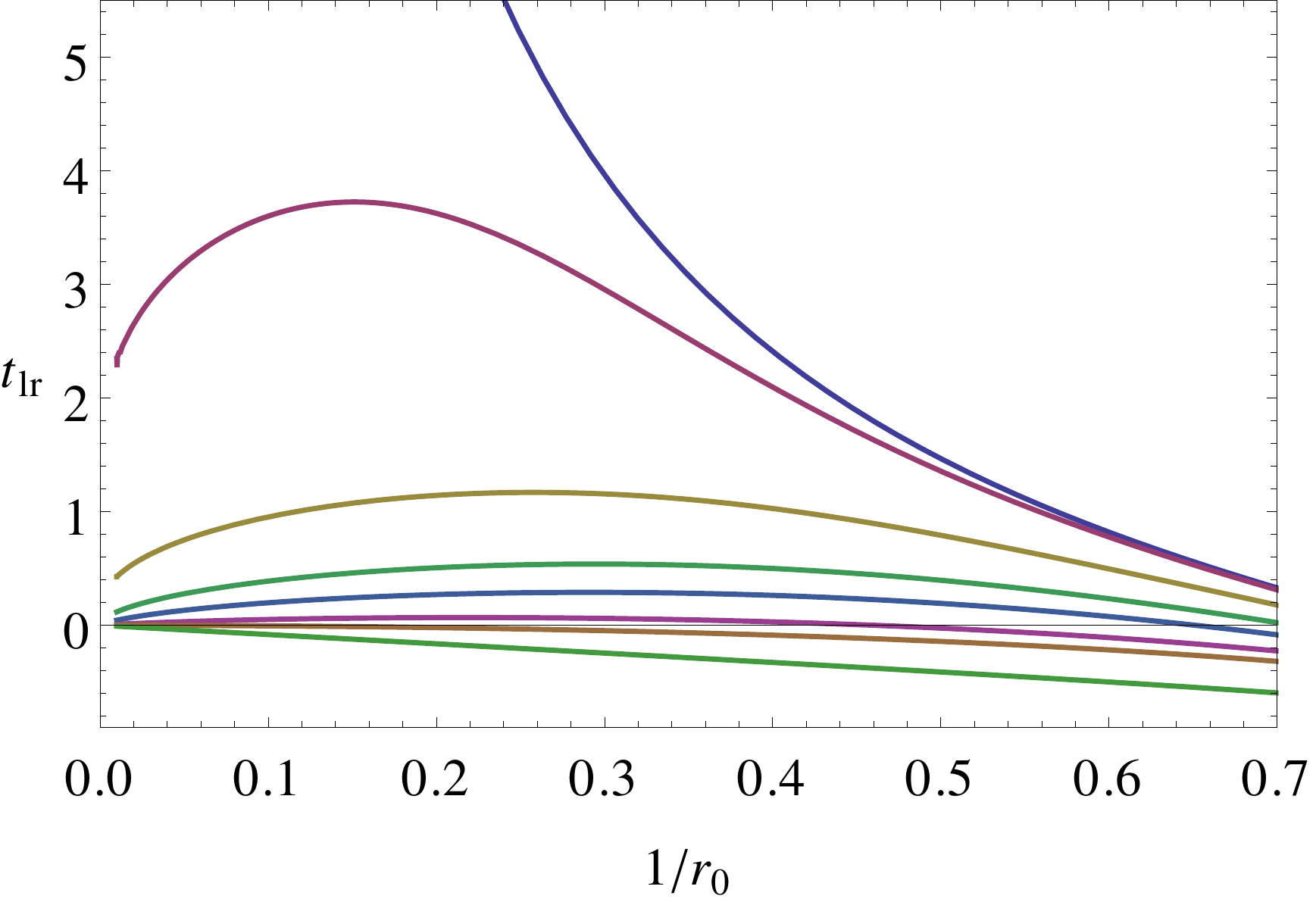}
\caption{\label{fig:lastray} The time, when the last ray of light to reach the shell is sent from the boundary. The horizontal axis in the plot is the initial location of the shell $1/r_0$, while the different curves correspond to $c=1/3$, $0.33$, $0.3$, $0.25$, $0.2$, $0.1$, $0$, $-1$ (from top to bottom). It is interesting to note that only the highest and two lowest curves corresponding to $1/3$, $0$ and $-1$ are monotonous: For $0 < c < 1/3$, there exists an `optimal' value of $r_0$, for which the collapse of the shell is the slowest.}
\end{center}
\end{figure}

While it takes an infinite amount of outside coordinate time $t_+$ for the shell to reach its Schwarzschild radius $r_h=1$, the corresponding values of the interior and proper times, $t_-$ and $\tau$, are finite. An important point for our following analysis is that there always exists a finite value of $t_+$, denoted $t_{\mathrm{lr}}$, after which no null geodesic starting from the boundary will reach the shell. The last light ray to reach the shell above the horizon is visualized as the dotted line in the Penrose diagram of fig.~\ref{fig:penrose}. Fig.~\ref{fig:lastray} on the other hand displays the dependence of this parameter on both $r_0$ and $c$, exhibiting a divergent behavior in the limit $r_0\to\infty$ and $c=1/3$ as discussed above.

{\em Continuity conditions across the shell.}
In order to study various entropy-like quantities (and eventually other observables), we would like to know how geodesics and other objects behave in the space-time containing a collapsing thin shell. On either side of this object, the problem reduces to one in pure AdS or AdS-Schwarzschild space-time, but in addition we need to know how geodesics and extremal surfaces can be continued past the shell, i.e.~how they refract at this location. 

What makes the matching of objects across the shell nontrivial is that both our time coordinate and metric are discontinuous there. This is, however, only an artefact of our choice of coordinates. In fact, one can explicitly construct a coordinate system, in which all components of the metric are continuous. In this coordinate system, all quantities whose EoMs do not involve derivatives of the metric higher than first order --- geodesics, surfaces, fields, etc.~--- are automatically continuous along with their first derivatives.

To construct the continuous coordinate system, we clearly need to replace $t$ and $r$ by new coordinates. To this end, we first determine the proper distance of a given point to the shell along a geodesic normal to it by solving the equations
\begin{equation}
u^\mu = \frac{dx^\mu}{d\lambda} \quad\text{and}\quad \nabla_u u =0 \quad\text{such that}\quad u\big|_{\lambda=0} = n \; ,
\label{eq:geodesiceq}
\end{equation}
where $n$ is a normal vector of the shell, and then take the parameter $\lambda$ as our new radial coordinate. As our time coordinate, we on the other hand choose the proper time of the shell $\tau$ along the above geodesic. That is, given a point in the $(t,r)$-coordinates, we find the corresponding values of $\tau$ and $\lambda$ by determining the spacelike geodesic that is normal to the shell and intersects this point; $\lambda$ is then the proper distance along this geodesic, and $\tau$ the proper time of the shell at its intersection point with the geodesic.

Applying the above coordinate transformation, our metric obtains the form
\begin{equation}
ds^2=-G(\lambda,\tau)^2d\tau^2+d\lambda^2+r(\lambda,\tau)^2d\textbf{x}^2,
\end{equation}
where the function
\begin{equation}
G(\lambda,\tau)=\sqrt{\frac{f(r)+\dot{r}_s^2}{f(r_s)+\dot{r}_s^2}}+\sqrt{f(r)+\dot{r}_s^2}\int_0^\lambda \frac{d\lambda\,\ddot{r}_s}{f(r)+\dot{r}_s^2}
\end{equation}
is manifestly continuous at the shell. Here, $r$ denotes the inverted function $r=r(\lambda,\tau)$, while $f$ refers to $f_+$ outside the shell ($\lambda>0$) and to $f_-$ inside it ($\lambda<0$). At the location of the shell, we in particular have $G(0,\tau)=1$.

In practical applications, we often wish to work in the original $(t,r)$ coordinates, and thus need to be able to phrase matching conditions in terms of them. Assuming that the object in question is parametrized by a coordinate $x$ that is continuous at the shell, we obtain
\begin{align}
\Big(\frac{dt}{dx}\Big)_+&=\frac{\dot{r}_s}{f_+f_-}(\beta_--\beta_+)\Big(\frac{dr}{dx}\Big)_-\\ \nonumber &\qquad+\frac{1}{f_+}(\beta_+\beta_--\dot{r}_s^2)
\Big(\frac{dt}{dx}\Big)_-
\\
\Big(\frac{dr}{dx}\Big)_+&=\frac{1}{f_-}(\beta_+\beta_--\dot{r}_s^2)
\Big(\frac{dr}{dx}\Big)_-\\ \nonumber &\qquad+\dot{r}_s(\beta_--\beta_+)\Big(\frac{dt}{dx}\Big)_-,
\end{align}
where we have defined $\beta_{\pm}\equiv\sqrt{f_{\pm}+\dot{r}_s^2}$, and it is understood that all terms are evaluated at the location of the shell.

\begin{figure}[t]
\begin{center}
\includegraphics[scale=0.45]{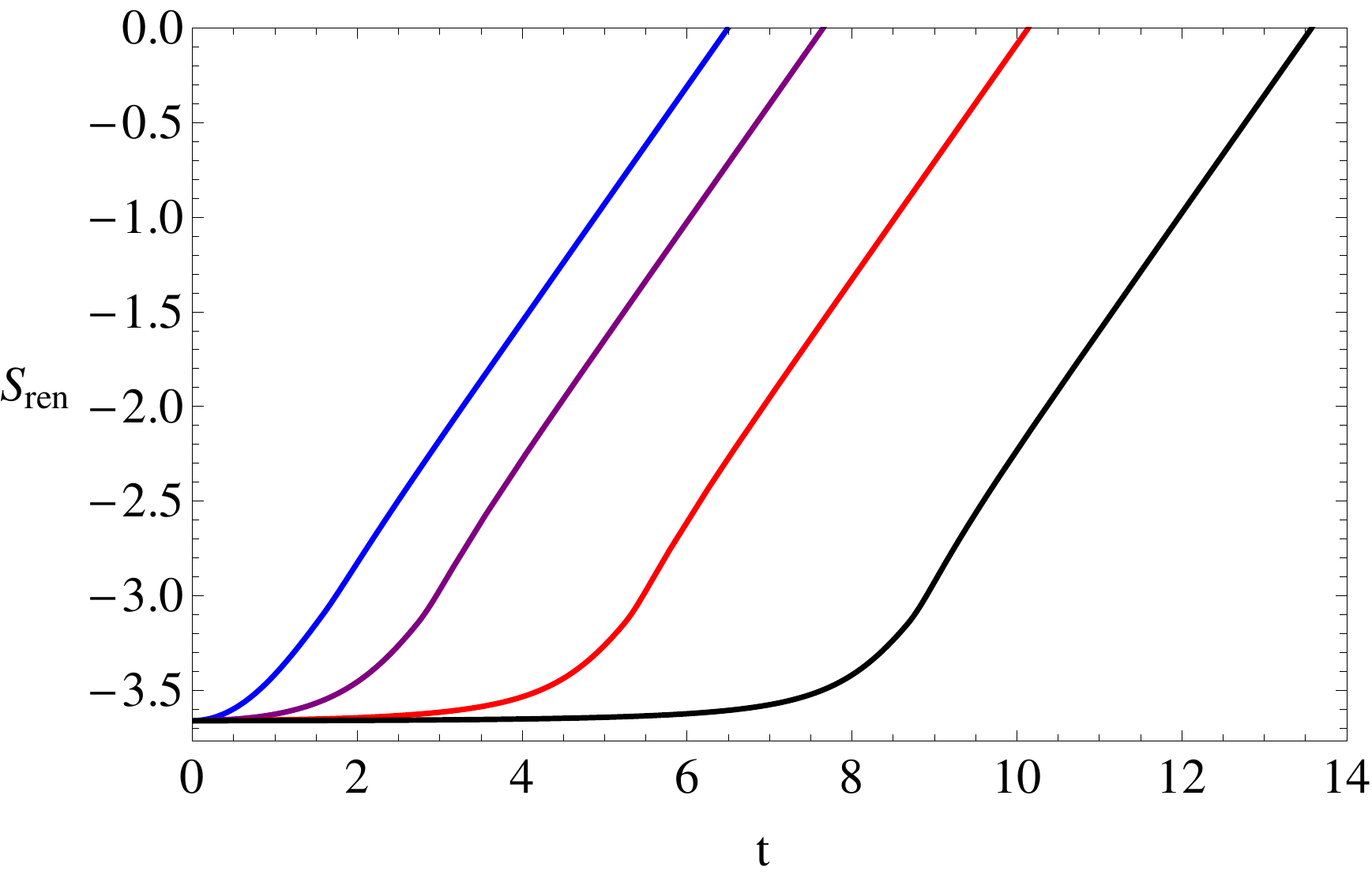}
\caption{\label{fig:stripc} Time evolution of the HEE for a strip-like boundary region of depth $R=8$ and for shells starting from $r_0=5$ and described by EoS parameters $c=0$, 0.3, 0.33 and 1/3 (from left to right).}
\end{center}
\end{figure}

{\em The entropies.}
To probe equilibration in the dual field theory, we use two observables for which the machinery developed above turns out to be highly useful. We divide the system to a subsystem $A$ and its exterior, and quantify the stage of thermalization by studying the time evolution of the density matrix of the subsystem towards its maximally mixed thermal limit. In principle, such a calculation can be performed in several ways, but the most convenient approach for holographic applications turns out to be via the von Neumann --- or entanglement --- entropy
\beq
S=-\textrm{tr}\rho_A\log \rho_A\, ,
\eeq
where $\rho_A$ is the density matrix of $A$. The holographic dual of this quantity (HEE) is given by the extremal surface prescription \cite{Ryu:2006bv,Hubeny:2007xt} as
\beq
S=\frac{A_\text{ext}}{4G_N} \, ,
\eeq
where $A_\text{ext}$ denotes the area of an extremal surface ending at the boundary of $A$.

The second quantity we consider is the causal holographic information (CHI)
\beq
\chi=\frac{A_\text{causal}}{4G_N},
\eeq
where $A_\text{causal}$ denotes the area of the Rindler horizon of a causal diamond on the boundary, centered at a constant time slice. The field theory interpretation of this quantity is not yet settled. A particularly intriguing suggestion was made in \cite{Kelly:2013aja}, where it was conjectured that $\chi$ might be dual to a particular coarse grained entropy, the entropy of a maximally mixed state, in which the values of all one-point functions of local gauge invariant operators are fixed.

The HEE is obtained by solving the Euler-Lagrange equations derived from the extremization of the area functional
\beq
A_\text{ext}=\int d^3\sigma\sqrt{\det_{ab}g_{\mu\nu}\frac{\partial x^{\mu}}{\partial\sigma^a}\frac{\partial x^{\nu}}{\partial\sigma^b}},
\eeq
in which the subsystem $A$ is typically taken to have the shape of either a sphere or an infinite strip. Extremal surfaces in AdS and AdS-Schwarzschild space-times have been studied before (see e.g.~\cite{Ryu:2006bv,Hubeny:2012wa,Hubeny:2013hz}), and the only subtleties in our case are related to the presence of the collapsing shell, i.e.~the matching conditions of the extremal surface at this time-dependent location. Below, we proceed by first solving the shell EoM numerically, and then finding the extremal surface. 

\begin{figure}[t]
\begin{center}
\includegraphics[scale=0.45]{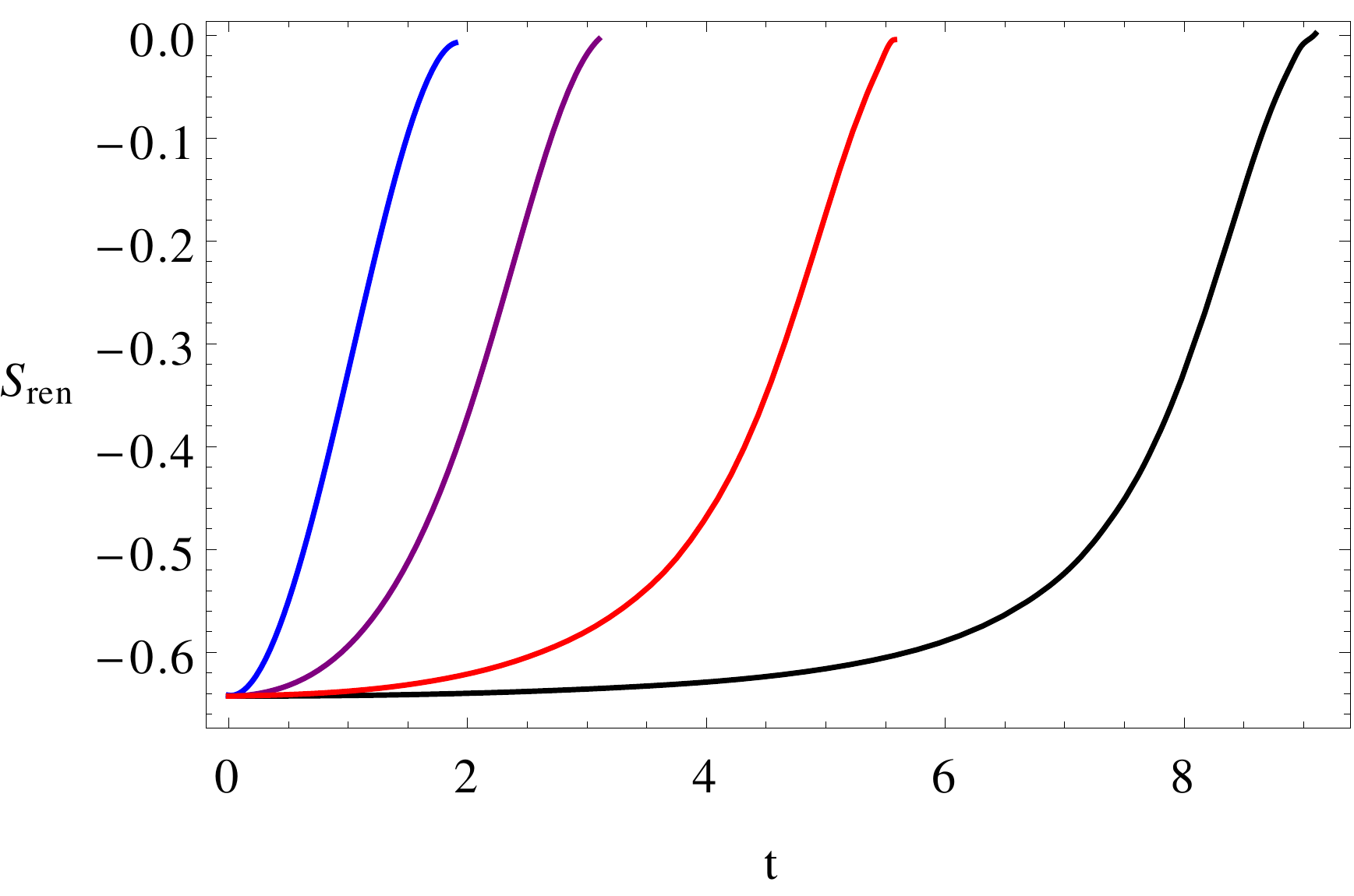}
\caption{\label{fig:sph1} Time evolution of the HEE for a sphere with radius $R=2$ and a shell starting off at $r_0=5$. The different curves correspond to $c=(0,0.3,0.33,1/3)$ (from left to right).}
\end{center}
\end{figure}

\begin{figure}[t]
\begin{center}
\includegraphics[scale=0.45]{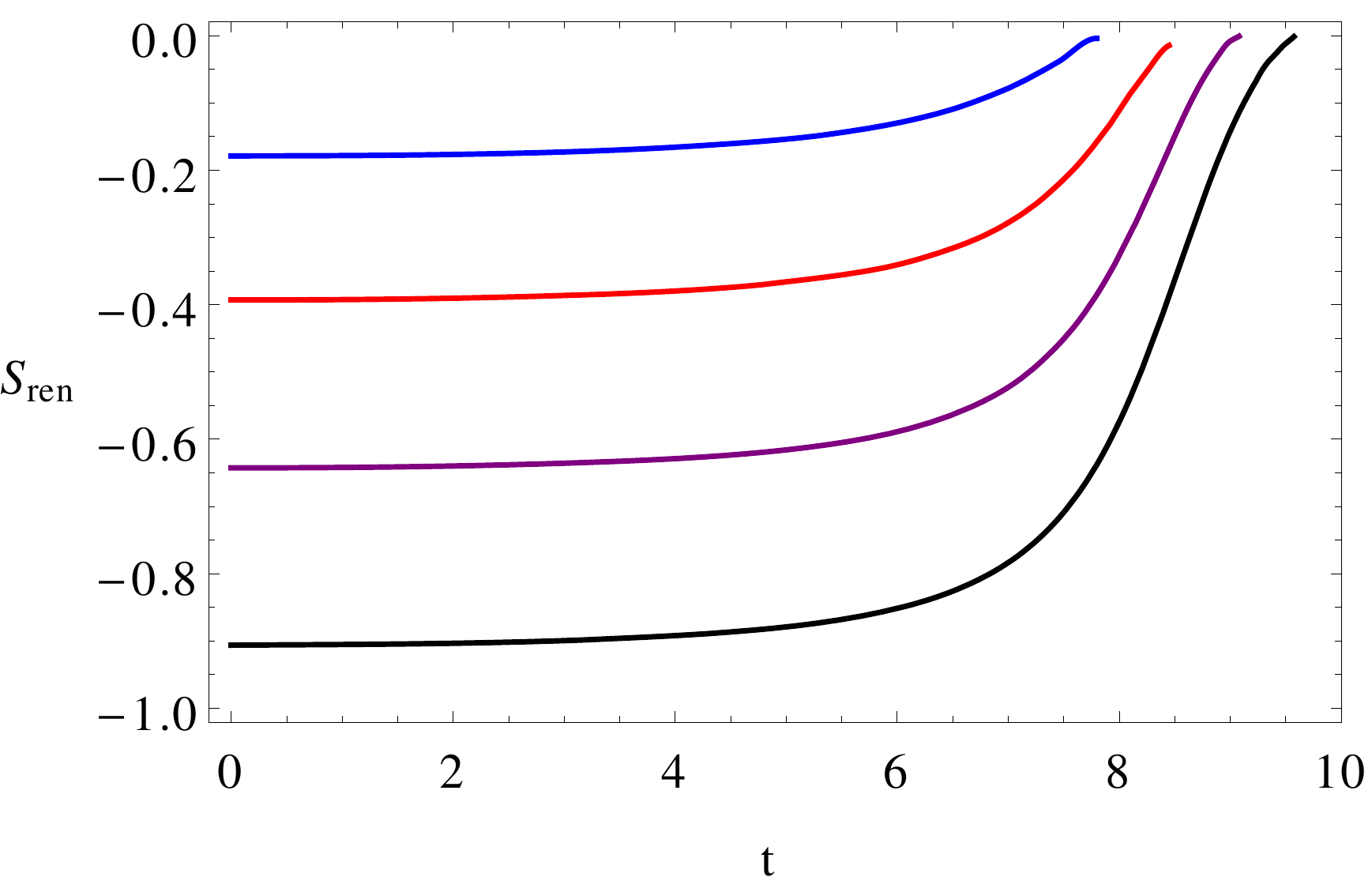}
\caption{\label{fig:sph2} Same as in fig.~\ref{fig:sph1}, but for $c=1/3$, $r_0=5$ and $R=1,\;1.5,\;2,\;2.5$ (from top to bottom).}
\end{center}
\end{figure}

As we start the shell off at some finite $r=r_0$, it is expected that sufficiently small spatial regions have a thermal value of the HEE right from the beginning. In the following, we will thus consider the more interesting situation of larger regions, in which the HEE starts from a non-thermal value and the extremal surface passes through the shell. In these cases, we find that at early times the behavior of the HEE is consistent with the form
\beq
S(t)=S(0)+\gamma t^2+...,
\eeq
similar to the Vaidya result except for the nonzero offset $S(0)$. For the case of a strip region, a simple analytic calculation gives in the limit of a large boundary region
\begin{eqnarray}
\gamma_\text{strip} &=& -\frac{A_{\partial \! A}}{8G_N}  \frac{ar_0\sqrt{f_+(r_0)}\left[r_0-\sqrt{f_+(r_0)}\right]}{1+\frac{1}{2}r_0 a\left[r_0-\sqrt{f_+(r_0)}\right]^2},\label{eq:gamma}
\end{eqnarray}
where $A_{\partial \! A}$ denotes the surface area of the boundary strip and $a=\frac{d^2 r_s}{d\tau^2}|_{t=0}$ the initial acceleration of the shell. For the case of a spherical region, our numerical results suggest the identification
\beq
\gamma_\text{sphere}=\frac{3}{2}\gamma_\text{strip},
\eeq
where $A_{\partial \! A}$ is now understood as the area of the boundary sphere. These results are shown in figs.~\ref{fig:stripc}--\ref{fig:sph2}, where the subscript `ren' stands for the fact that we have normalized the quantity by subtracting from it its final equilibrium value and then divided the remainder by $A_{\partial A} /4 G_N$. Also, we should note that by `early times' we mean here $t\ll 1/\sqrt{-a}$.

At later times, we find that the HEE exhibits linear growth for both boundary regions, taking the form
\beq
S(t)\approx \, \textrm{constant}+A_{\partial \!A} s_\text{eq} v_E\,  t,\label{eq:linearpart}
\eeq
where $v_E=\sqrt{2}/3^{3/4}\approx 0.6204$ and $s_\text{eq}=1/(4G_N)$ stands for the entropy density of the final equilibrium state. This is exactly what was found in the Vaidya case \cite{Liu:2013qca}, and constitutes one of our main results. The reason for the behavior is simple: In the case of large boundary regions, the intermediate time behavior of the HEE is known to be dominated by a critical extremal surface that goes through the shell and inside the black hole (apparent) horizon \cite{Liu:2013qca,Hartman:2013qma}. Most of the area of the surface arises from the region inside the horizon. The only difference between our setup and that of \cite{Liu:2013qca} is that the trajectory of our shell is not lightlike and the matching conditions are somewhat different; this, however, turns out to be unimportant in the regime of linear growth. In the strip region example of fig.~\ref{fig:stripc}, we indeed observe the emergence of a long linear part with a slope well approximated by eq.~(\ref{eq:linearpart}). The late time behavior of the quantity is also found to be consistent with the Vaidya results, and it seems likely that the corresponding analysis by Liu and Suh could be generalized to our case. We have, however, not investigated this issue. 

\begin{figure}[t]
\begin{center}
\includegraphics[scale=0.45]{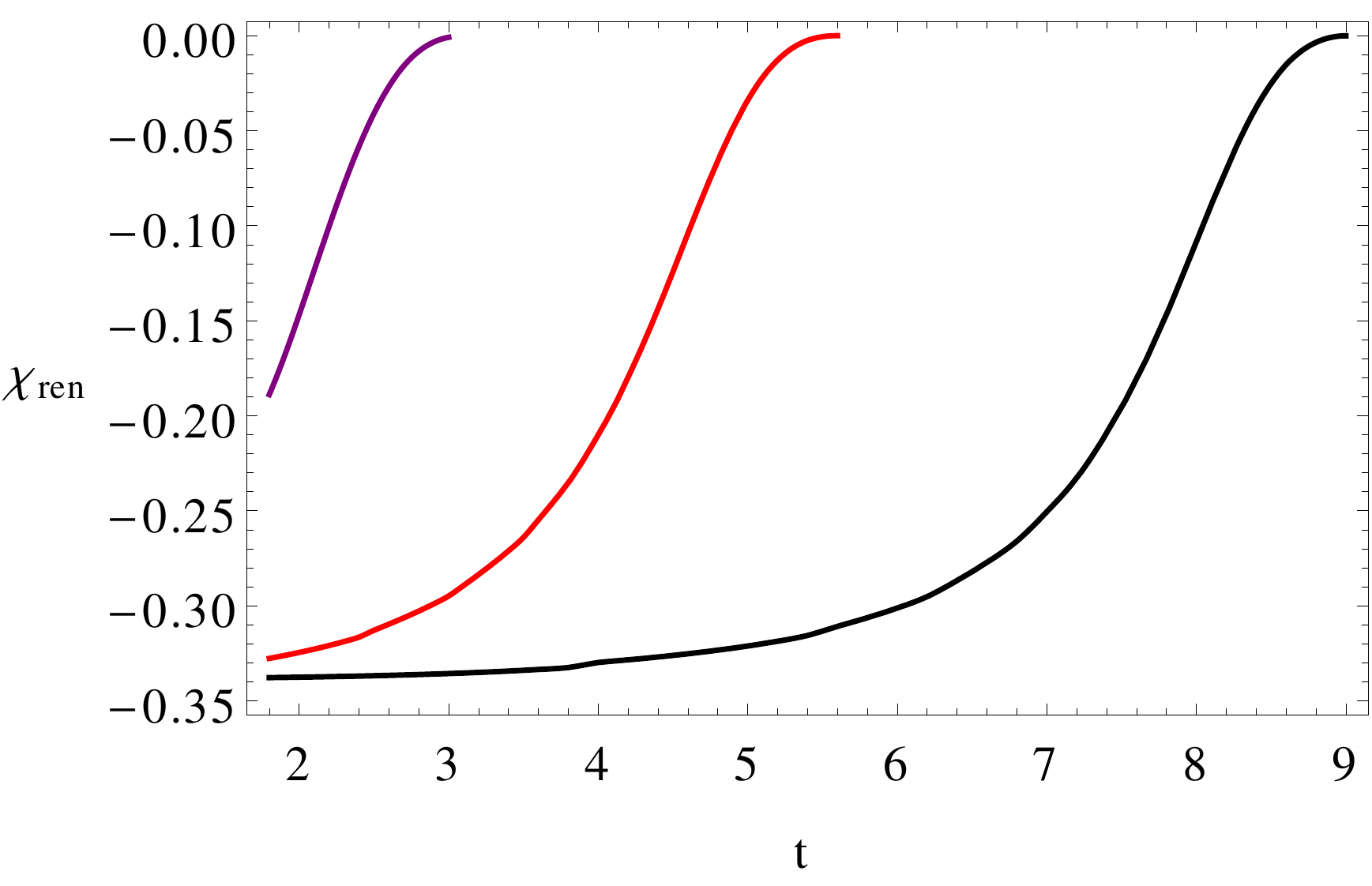}
\caption{\label{fig:CHI1} The CHI, evaluated for $R=2$ and $r_0=5$, with $c$ taking the values 0.3, 0.33 and 1/3 (from left to right).}
 \end{center}
\end{figure}

Finally, we note that for the HEE the thermalization time is seen to depend on both the shape of the boundary region and the trajectory of the shell. For large boundary regions, we can derive an approximative relation
\beq
t_\text{therm}\approx t_\text{lr}+\alpha R,
\eeq
where $t_\text{lr}$ is the boundary time of the last ray defined above and depends only on the shell trajectory. The coefficient $\alpha$ on the other hand carries dependence on the shape of the boundary region, but is otherwise universal: For a strip, one gets $\alpha=1/(2v_E)$, and for a sphere $\alpha=\sqrt{3/2}$.

Next, we move on to consider the CHI. It is obtained from the Rindler horizon of a boundary causal diamond, which is constructed by considering null geodesics in the space-time and finding the maximal bulk region that is in causal contact with the boundary causal diamond of size $R$ centered at time $t$. The CHI is then given by the minimal surface area of this region. Null geodesics in AdS and AdS-Schwarzschild space-time have been considered in several earlier works (see e.g. \cite{Hubeny:2012wa, Hubeny:2013hz}), and once again the only complications in our case originate from the matching conditions and the specific trajectory of the shell. In the following, we briefly summarize our numerical results for this quantity. 

\begin{figure}[t]
\begin{center}
\includegraphics[scale=0.45]{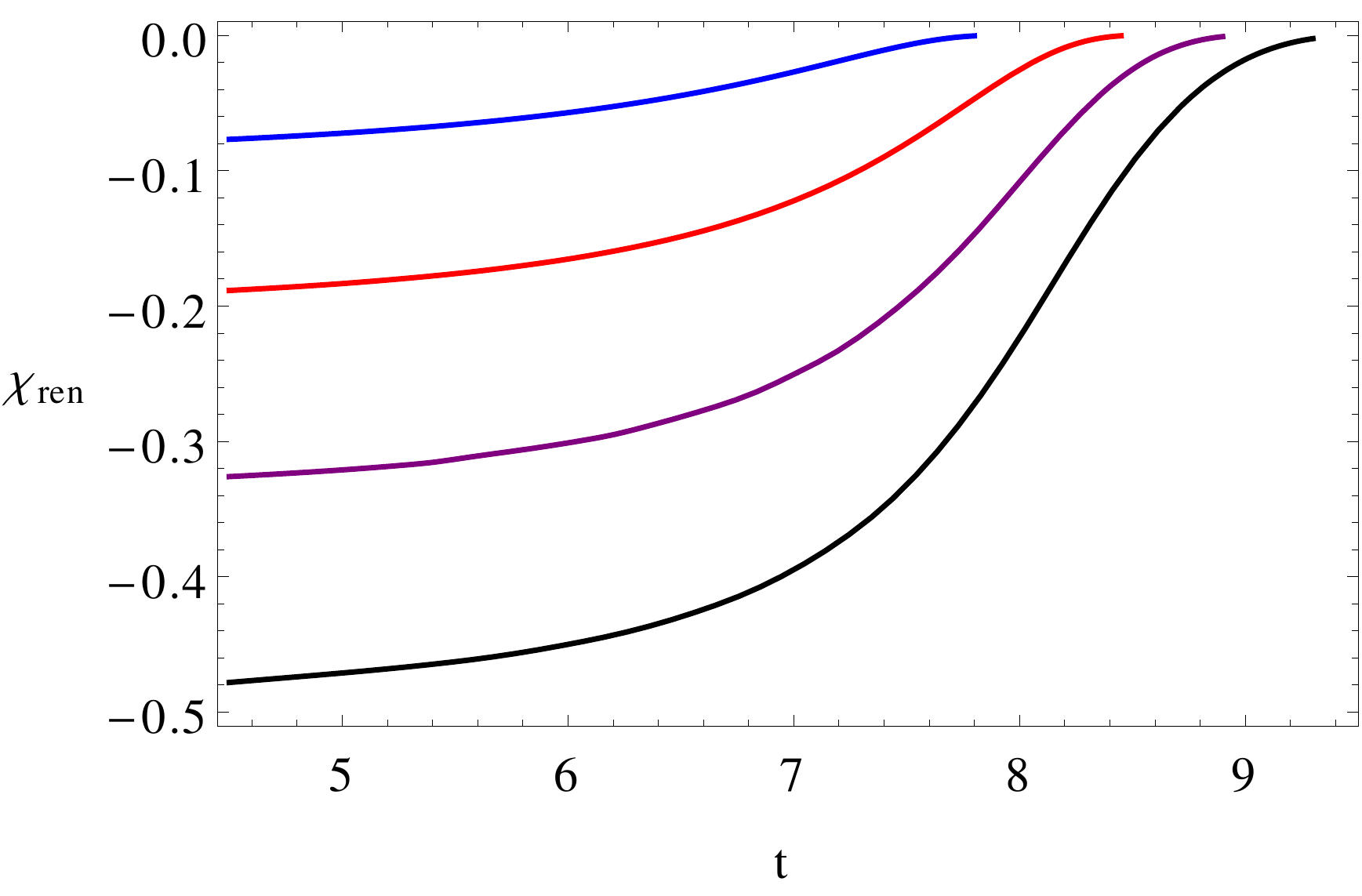}
\caption{\label{fig:CHI2} The CHI for $c=1/3$ and $r_0=5$,  with $R$ taking the values 1, 1.5, 2, 2.5 (from top to bottom).}
 \end{center}
\end{figure}

In Vaidya space-time, the CHI relaxes to a constant thermal value at the time $t=R$, when the past tip of the boundary causal diamond passes the time $t=0$, at which the shell is released. This happens because the shell follows a null trajectory, which is no longer the case in our calculation; thus, for us the CHI turns out to have non-trivial time dependence even after $t=R$, which is what we focus on in figs.~\ref{fig:CHI1} and \ref{fig:CHI2}. From these figures, we observe that the quantity monotonically increases in time and equilibrates qualitatively very similarly to the HEE.

The thermalization pattern of the CHI can be easily understood. Since any light ray sent after the time $t_\text{lr}$ will catch the shell only behind its event horizon, the CHI thermalizes when the past tip of the boundary causal diamond passes $t=t_\text{lr}$. Thus, at large $R$  the thermalization time of the CHI is given by the simple formula
\beq
t_\text{therm}=t_\text{lr}+R.
\eeq

{\em Conclusions.}
In the paper at hand, we have studied the equilibration dynamics of a strongly coupled quantum field theory in a specific holographic setup dual to a thin shell of matter starting from rest at a given radial coordinate in the AdS$_5$ bulk and collapsing along a timelike trajectory to form a black hole. In our calculation, we concentrated on the dynamics of shells with different equations of state and the matching conditions of various objects at their location, applying the results to the evaluation of two entropy-like quantities in this dynamical background. 

In many ways, our calculation can be viewed as a continuation and generalization of the earlier works of \cite{Liu:2013qca,Baron:2012fv} (see also
\cite{Alishahiha:2014cwa,Fonda:2014ula} for further generalizations of \cite{Liu:2013qca}). While ref.~\cite{Liu:2013qca} considered lightlike shells, our shells are timelike and released from rest at a finite radial location. This allows us to study a broader class of collapsing spacetimes, from which we can obtain the Vaidya results as a limiting case, valid at late times for generic shell trajectories. Typically, we found the equilibration process to proceed at a somewhat slower pace than in the Vaidya case, with the difference being the most pronounced for a conformal shell. At early times, we furthermore found the quadratic time dependence of the HEE to be sensitive to the trajectory of the shell, and in fact proportional to its initial acceleration. This is in stark contrast to the Vaidya limit, where the early time increase of the HEE is known to be universal and depend only on the energy density of the non-equilibrium state and the area of the boundary entangling surface.

Our paper is not the first one to study the gravitational collapse of shells with timelike trajectories; this question was already addressed in \cite{Baron:2012fv}. We were, however, the first ones to work out the matching conditions of geodesics and other objects at the position of the shell, allowing us to obtain more analytic control over the time-evolution of a class of physical observables. For example, we were able to analytically determine the early time dynamics of the HEE, explaining some of the numerical findings of \cite{Baron:2012fv}, and to find a linear regime in the time evolution of the HEE, as proposed by \cite{Liu:2013qca}. Finally, we expanded the study of \cite{Baron:2012fv} to different shapes of the boundary entangling region and to the thermalization of the CHI, and furthermore presented a systematic study of the effects of the initial position of the shell on the results.

The work we have reported in this paper marks only one modest step towards a more complete understanding of the equilibration dynamics taking place in strongly coupled field theory. In particular, while we have successfully generalized previous works on the collapsing shell scenario to include different initial states and shell equations of state, we have barely touched upon the field theoretical interpretation of these choices. This marks one of the main questions we are planning to address in a later publication that will in addition contain a much more detailed account of the calculations presented in this paper.

{\em Acknowledgments.}
We are grateful to Janne Alanen for helpful discussions as well as collaboration in the early stages of this work. In addition, we thank Rolf Baier, Daniel Grumiller, Keijo Kajantie, Esko Keski-Vakkuri, Anton Rebhan, and Bin Wu for useful discussions. H.N., S.S., O.T.~and A.V.~were financially supported by the Sofja Kovalevskaja program of the Alexander von Humboldt Foundation, S.S.~by the FWF projects Y435-N16 and P26328, O.T.~by the Dutch Foundation for Fundamental Research of Matter (FOM), and A.V.~by the Academy of Finland, grant \# 266185. The research of VK was supported by the European Research Council under the European Union's Seventh Framework Programme (ERC Grant agreement 307955). V.K., S.S.~and O.T.~in addition acknowledge the hospitality of the Helsinki Institute of Physics, and S.S.~and O.T.~that of the University of Oxford. V.K.~and A.V.~also thank the Institute of Theoretical Physics at the Technical University of Vienna and the ESF network \textit{Holographic methods for strongly coupled systems} (HoloGrav) for financial support.


\end{document}